\documentclass[aps,pre,twocolumn,showkeys,superscriptaddress]{revtex4}
\usepackage{epsfig}
\usepackage{times}
\begin{document}

\title{Multiple firing coherence resonances in excitatory and inhibitory coupled neurons}

\author{Qingyun Wang}
\email{nmqingyun@163.com}
\affiliation{Department of Dynamics and Control, Beihang University, Beijing, China}
\author{Honghui Zhang}
\affiliation{Department of Dynamics and Control, Beihang University, Beijing, China}
\author{Matja\v{z} Perc}
\email{matjaz.perc@uni-mb.si}
\affiliation{Department of Physics, Faculty of Natural Sciences and Mathematics, University of Maribor, Slovenia}
\author{Guanrong Chen}
\affiliation{Department of Electronic Engineering, City University of Hong
Kong, Hong Kong SAR, China}

\begin{abstract}
The impact of inhibitory and excitatory synapses in delay-coupled Hodgkin--Huxley neurons that are driven by noise is studied. If both synaptic types are used for coupling, appropriately tuned delays in the inhibition feedback induce multiple firing coherence resonances at sufficiently strong coupling strengths, thus giving rise to tongues of coherency in the corresponding delay-strength parameter plane. If only inhibitory synapses are used, however, appropriately tuned delays also give rise to multiresonant responses, yet the successive delays warranting an optimal coherence of excitations obey different relations with regards to the inherent time scales of neuronal dynamics. This leads to denser coherence resonance patterns in the delay-strength parameter plane. The robustness of these findings to the introduction of delay in the excitatory feedback, to noise, and to the number of coupled neurons is determined. Mechanisms underlying our observations are revealed, and it is suggested that the regularity of spiking across neuronal networks can be optimized in an unexpectedly rich variety of ways, depending on the type of coupling and the duration of delays.
\end{abstract}

\keywords{coherence resonance, synaptic coupling, information transmission delay, regularity of spiking, time scales}
\maketitle

\section{Introduction}
Neurophysiological studies have revealed the existence of accurately timed patterns of spikes by a variety of cognitive and motoric tasks \cite{lestienne_br87, abeles_jn93, riehle_s97, oram_jn99, johansson_nn04, pipa_nc07}. The timing of these spikes, or neuronal firings, is accurate to within the millisecond range, which poses great challenges with regards to the identification of mechanisms that would be able to ensure such precision. Following their initial observation in the cortex of monkeys \cite{lestienne_br87, abeles_jn93}, the precisely timed spikes have been reported and investigated for motor functions \cite{riehle_s97}, the neuronal response of visual systems \cite{oram_jn99}, and the complex spatial fingertip events \cite{johansson_nn04}, to name but a few examples. Not surprisingly, synchronized, precisely timed firings can be observed at virtually all neuronal processing levels, including the retina \cite{neuenschwander_n96}, the lateral geniculate nucleus \cite{branco_ns98}, and the cortex \cite{gray_n89, fries_pnas97}.

Since it is well known that noise can play a constructive role in different types of nonlinear dynamical systems, which arguably describe also neuronal dynamics \cite{hodgkin_jp52}, this opens the possibility of exploiting such mechanisms for explaining, or at least supporting, the aforementioned precision of neuronal firings. Stochastic resonance \cite{benzi_jpa81, nicolis_tel81, gammaitoni_rmp98} and coherence resonance \cite{hu_prl93, longtin_pre97, pikovsky_prl97} are amongst the most prominent examples by means of which noise of appropriate intensity is able either to enhance the detection of weak deterministic signals \cite{hanggi_cpc02} or evoke coherent response in nonlinear dynamical systems in the absence of any deterministic inputs. The potential benefits of noise range from ice ages to crayfish and SQUIDs \cite{wiesenfeld_n95}, to neural systems, as most recently reviewed in \cite{mcdonnell_nrn11}.

Following initial advances on individual dynamical systems, the focus begun shifting to spatially extended systems \cite{saques_rmp07}, especially also to such with complex networks describing connections between the individual units \cite{albert_rmp02, boccaletti_pr06}. For example, coherence resonance on a small world network was investigated in \cite{kwon_pla02}, while array-enhanced resonances were reported in \cite{zhou_pre03}. Moreover, spatial coherence resonance was observed first near pattern-forming instabilities \cite{carrillo_epl04}, and latter also in excitable media \cite{perc_pre05}. Excitable systems in general proved to be very susceptible to a multitude of noise-induced phenomena, as reviewed comprehensively in \cite{lindner_pr04}. Adding spatial degrees of freedom, along with the possibilities for introducing other sources of heterogeneity, lead to the discovery of very interesting and quite exotic phenomena, such as the ghost resonance \cite{balenzuela_bs07}, and double as well as multiple stochastic \cite{zaikin_prl03, wang_chaos09, gan_cpb10, zeng_epjd11} and coherence \cite{horikawa_pre01, kreuz_prl06, baohua_cpb09, xiu_chaos11} resonances.

For neural systems, a wealth of interesting and new phenomena was made observable by integrating realistic features of neuronal dynamics into the studied models. Information transmission delays or synaptic delays, for example, are inherent to the nervous
system because of the finite speed at which action potentials propagate across neuron axons, and due to time lapses occurring at both dendritic and synaptic processing \cite{kandel_91}. Following seminal works examining the impact of delays on excitable and other dynamical systems \cite{fatihcan_prl03, sethia_pla07, sethia_prl08}, the stability and attainability of synchronous oscillations \cite{rossoni_pre05, wang_epl08, wang_pre09a} and the role of delays in shaping spatiotemporal dynamics of neuronal activity \cite{roxin_prl05} were investigated. Moreover, the role of delays in coupled Hodgkin-Huxley neurons was also investigated for the phenomenon of coherence resonance, and it was reported that properly tuned delays can lead to the occurrence of multiple resonances \cite{gong_bs11, hao_nc11}.

In this letter, we extend the scope of coherence resonance in models of neuronal dynamics by considering besides synaptic delays also different types of synaptic coupling. While the role of chemical synapses in coupled neurons with noise has been investigated in \cite{balenzuela_pre05}, and although the general dynamics of sparsely connected networks of excitatory and inhibitory spiking neurons is known \cite{brunel_jcn00}, our approach, joining these distinctive features of neuronal dynamics (synaptic delays, different types of synaptic coupling, and noise), allows for the identification of new ways by means of which the coherence, and thus the accuracy of neuronal firings, can be improved. Most interestingly, we report the occurrence of multiple coherence resonance patterns in the corresponding delay-strength parameter plane when either inhibitory and excitatory or only inhibitory synapses are used for coupling. The details of these multiple firing coherence resonances, and in particular the conditions at which they occur, however, depend significantly on the type of coupling. Reported results suggest that characteristic time scales related to the information transmission and inhibition in neuronal networks may interplay in intricate ways, and by doing so give rise to new mechanisms for optimizing spiking regularity.

The remainder of this letter is organized as follows. In the next section we describe the model, then we present the main results separately for the two coupling scenarios, while lastly we summarize our findings and discuss their potential implications.

\section{Model definition}
For simplicity, we consider two Hodgkin--Huxley neurons \cite{hodgkin_jp52} that are coupled by inhibitory and/or excitatory synapses. Equations describing the dynamics are:
\begin{eqnarray}
C \frac{{\rm d}V_{_{i}}}{{\rm d}t}&=&-g_{Na}m^3h(V_{i}-V_{Na})-g_{L}(V_{i}-V_{L})\nonumber \\
& &-g_{K}X_{K}n^4(V_{i}-V_{K})+I+\sigma \xi_{i}(t)+I^{i,j}_{syn},\\
\frac{{\rm d}m_{i}}{{\rm d}t}&=&\alpha_{m_{i}}(1-m_{i})-\beta_{m_{_{i}}}{m_{i}},\\
\frac{{\rm d}h_{i}}{{\rm d}t}&=&\alpha_{h_{i}}(1-h_{i})-\beta_{h_{i}}{h_{i}},\\
\frac{{\rm d}n_{i}}{{\rm d}t}&=&\alpha_{n_{i}}(1-n_{i})-\beta_{n_{i}}{n_{i}},
\end{eqnarray}
where $V_{i}$ is the transmembrane potential of the $i$-th neuron. Moreover, $m_{i}$, $h_{i}$ and $n_{i}$ are the gating variables, where the voltage-dependent opening and closing rates are:
\begin{eqnarray}
\alpha_{m_{i}}&=&\frac{0.1(V_{i}+10)}{1-{\rm exp}[-\frac{(V_{i}+40)}{10}]}, \\
\beta_{m_{i}}&=&4 {\rm exp}\left[-\frac{(V_{i}+65)}{18} \right], \\
\alpha_{h_{i}}&=&0.07 {\rm exp}\left[-\frac{(V_{i}+65)}{20}\right], \\
\beta_{h_{i}}&=&\left\{1+{\rm exp}\left[-\frac{(V_{i}+35)}{10}\right]\right\}^{-1}, \\
\alpha_{n_{i}}&=&\frac{0.01(V_{i}+55)}{1-{\rm exp}[-\frac{(V_{i}+55)}{10}]}, \\
\beta_{n_{i}}&=&0.125 {\rm exp}\left[-\frac{(V_{i}+65)}{80}\right],
\end{eqnarray}
The membrane capacity is $C=1$ ($\mu$F/cm$^{2}$), and $g_{Na}=120$ $\mu$F/cm$^{2}$, $g_{K}=36$ $\mu$F/cm$^{2}$ and $g_{L}=0.3$ $\mu$F/cm$^{2}$ are the maximal sodium, potassium and leakage conductances, respectively. The corresponding reversal potentials are $V_{Na}=50$ mV, $V_{K}=-77$ mV and $V_{L}=-54.4$ mV. Using these parameter values, a single Hodgkin--Huxley neuron has a subcritical Hopf bifurcation at the external current $I=I_1=9.8{\rm \mu A / cm^2}$. Between $I=I_2= 6.2{\rm \mu A/cm^2}$ and $I_1$ stable limit cycles coexist with stable steady states, whereas for $I < I_2$ ($I > I_1$) excitable steady states (limit cycles) are the only stable solutions. If $I>155 {\rm \mu A / cm^2}$, on the other hand, the oscillations vanish by means of a supercritical Hopf bifurcation. A more detailed bifurcation analysis of the Hodgkin--Huxley model was performed in \cite{wang_pre04, wang_DBDSB}. Here we are interested in the region $I < I_{2}$, where neurons are unable to fire spontaneously, i.e, remain forever quiescent in the absence of external stimuli. We thus set $I=6.1 {\rm \mu A / cm^2}$, so that both neurons are in an excitable steady state. Gaussian noise $\xi_{i}(t)$, having mean $<\xi_{i}(t)>=0$ and autocorrelation $<\xi_{i}(t) \xi_{j}(t')>= \delta_{ij} \delta(t-t')$, thus acts as the source of large-amplitude excitations, where $\sigma$ determines the noise intensity.

We consider two different coupling schemes. First, the two neurons are coupled in a hybrid way using inhibitory and excitatory synapses. The coupling terms in this case are:
\begin{eqnarray}
I^{1,2}_{syn}&=&-g_{exc}\frac{(V_{1}-V_{exc})}{(1+{\rm exp}\{-\lambda[V_{2}(t)-\Theta_{s}]\})},\\
I^{2,1}_{syn}&=&-g_{inh}\frac{(V_{2}-V_{inh})}{(1+{\rm exp}\{-\lambda[V_{1}(t-\tau)-\Theta_{s}]\})},
\end{eqnarray}
where the inhibitory feedback is delayed by $\tau$. Second, only inhibitory synapses are used for coupling, in which case the coupling becomes:
\begin{eqnarray}
I^{i,j}_{syn}&=&-g_{inh}\frac{(V_{i}-V_{inh})}{(1+{\rm exp}\{-\lambda[V_{j}(t-\tau)-\Theta_{s}]\})},
\end{eqnarray}
where the inhibitory feedback is again delayed by $\tau$, only that here this applies to both directions. In the above coupling terms $g_{inh(exc)}$ determines the strength of the synaptic conductance, i.e., the coupling strength, while $V_{inh}=-80$ mV and $V_{exc}=20$ mV are the reversal potentials for the inhibitory and the excitatory synapse, respectively. Moreover, $\Theta_{s}=0$ is the threshold, above which the postsynaptic neuron is affected by
the presynaptic one, and $\lambda=10$ is a constant rate for the onset of excitation or
inhibition. In what follows, we will investigate the impact of the delay $\tau$ and the coupling strength $g_{inh(exc)}$ on the occurrence of firing coherence resonance, and we will do so separately for the two described coupling schemes.

\section{Results}

\begin{figure}
\begin{center} \includegraphics[width = 8.4cm]{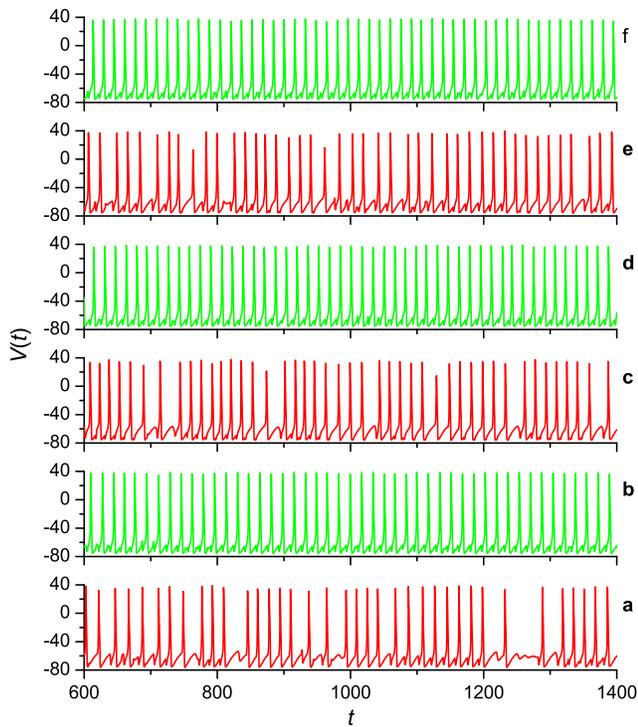}
\caption{\label{hybtime}Appropriately adjusted delays $\tau$ in the one-directional inhibition feedback enhance the regularity of spiking by hybrid coupling of the two neurons. Depicted are characteristic time courses of the transmembrane potential $V$ of the excitatory neuron for different values of $\tau$: (a) $0$, (b) $8.0$, (c) $20$, (d) $24$, (e) $35$ and (f) $40$. It can be observed that the regularity of spiking in panels (b), (d) and (f) (traces depicted green) is higher than in panels (a), (c) and (e) (traces depicted red). Other parameter values are: $g_{exc}=0.11$, $g_{inh}=1.0$ and $\sigma=1.5$.}
\end{center}
\end{figure}

We start by presenting the results as obtained with hybrid coupling, i.e., when excitatory and inhibitory synapses are used for connecting the two neurons. Figure~\ref{hybtime} features characteristic time courses of the transmembrane potential $V$ of the excitatory neuron, from where it can be observed at a glance that the coherence of excitations depends critically on the delay of the inhibitory feedback $\tau$. Importantly though, the relation between the coherency and the value of $\tau$ is not monotonous, but rather it is intermittent. That is to say, as $\tau$ increases the regularity is lost and regained intermittently as different values of $\tau$ come to determine the delay of inhibition. Time courses depicted green (panels b, d and f) exhibit more coherent spiking than time courses depicted red (panels a, c and e). This is characteristic for multiresonant phenomena, and in fact these observations can be made quantitatively more precise by introducing a coherence measure $C$ as follows. Let the sequence $t_{0} < t_1 < t_2 < \cdots< t_n$ denote the firing times of the considered neuron. From the sequence of $\{t_{k}\}$, the interspike intervals (ISI) are determined as $T_{k} = t_k -t_{k-1}(k = 1,2,\cdots,n)$. To characterize the coherence of the firings, the measure $C$ is defined as
\begin{equation}
C=\frac{\sqrt{<T^{2}_{k}>-<T_{k}>^{2}}}{<T_{k}>}.
\end{equation}
where $\langle \cdot \rangle$ is the time average. In particular, $C$ is the ratio of the standard deviation and the average of the interspike intervals, and it is indeed an excellent quantity for effectively determining the occurrence of coherence resonance from neuronal firing. From Eq.~(14) it follows that the more coherent
the firing, the smaller the value of $C$. We would also like to note that $C$ is the reciprocal of the coefficient of variation in a point process, which is widely used in the field of neuroscience \cite{koch_99}.

\begin{figure}
\begin{center} \includegraphics[width = 8.2236cm]{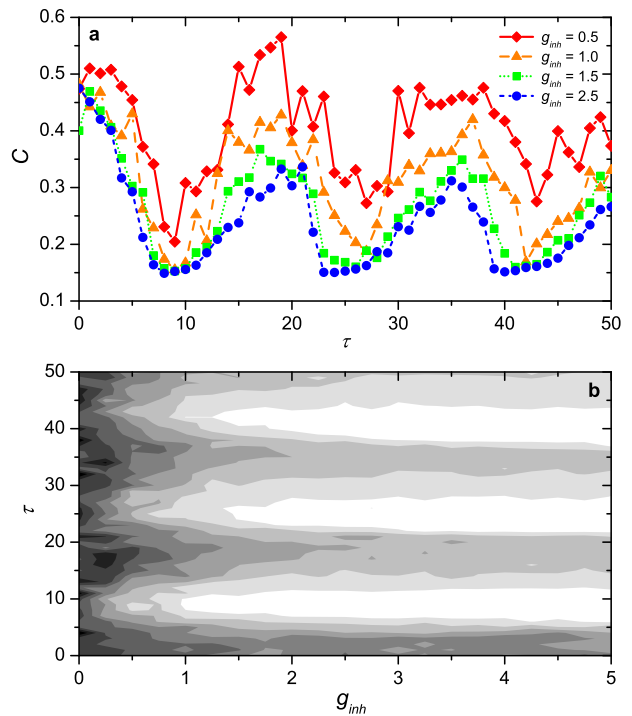}
\caption{\label{hybcr}Delay-induced multiresonances in case of hybrid coupling of the two neurons. Panel (a) shows the coherence measure $C$ in dependence on $\tau$ for different values of $g_{inh}$. It can be observed that the stronger the coupling the better expressed the recurrently appearing minima of $C$. Panel (b) features the contours of $C$ (white depicts minimal and black maximal values) on the corresponding delay-strength $\tau-g_{inh}$ parameter plane, where multiple tongues of coherency (white) emerge due to an interplay between the synaptic delay $\tau$ and the characteristic time scale of the two Hodgkin--Huxley neurons (as determined by the characteristic excitatory time $T_e$ and the complex conjugate part of the eigenvalues of the excitatory steady state). Other parameter values are: $\sigma=1.5$.}
\end{center}
\end{figure}

Using the introduced coherence measure $C$, we demonstrate in Fig.~\ref{hybcr} the occurrence of multiresonant behavior in dependence on $\tau$. Results presented in panel (a) indicate that $C$ has several minima in the considered interval of $\tau$, and that these are better pronounced, i.e., less susceptible to statistical deviations, for larger coupling strengths $g_{inh}$. In general, however, the dependence of $C$ on $g_{inh}$ is fairly insignificant, pointing towards the fact that in case of hybrid coupling the strength of the synaptic conductance of one type (e.g., the inhibitory type) has little impact if the other (e.g., the excitatory type) remains unchanged. The contours in panel (b) confirm this, as the tongues of coherency (white regions) simply shrink in width as $g_{inh}$ decreases, but otherwise do not alter the dependence of $C$ on the inhibition delay $\tau$. In many ways, these results are reminiscent of delay-induced multiple stochastic resonances that were previously reported for scale-free neuronal networks \cite{wang_chaos09}, and are indicative for an interplay between the time scales inherent to the system dynamics and the time scales introduced by means of the delay.

\begin{figure}
\begin{center} \includegraphics[width = 8.4cm]{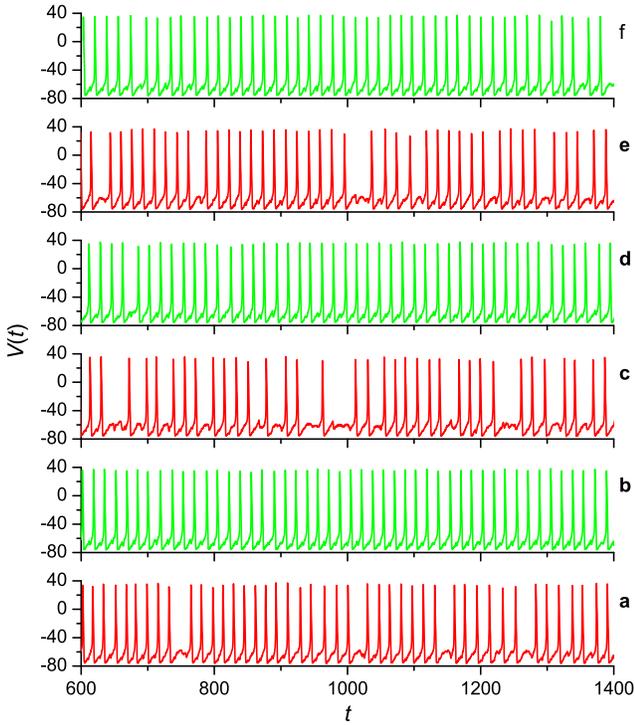}
\caption{\label{inhibtime}Appropriately adjusted delays $\tau$ in the bidirectional inhibition feedback enhance the regularity of spiking by inhibitory coupling of the two neurons. Depicted are characteristic time courses of the transmembrane potential $V$ of one neuron for different values of $\tau$: (a) $0$, (b) $2.0$, (c) $5.0$, (d) $11$, (e) $15$ and (f) $19$. As in Fig.~\ref{hybtime}, it can be observed that the regularity of spiking in panels (b), (d) and (f) (traces depicted green) is higher than in panels (a), (c) and (e) (traces depicted red). Other parameter values are: $g_{inh}=0.75$ and $\sigma=1.5$.}
\end{center}
\end{figure}

\begin{figure}
\begin{center} \includegraphics[width = 8.2236cm]{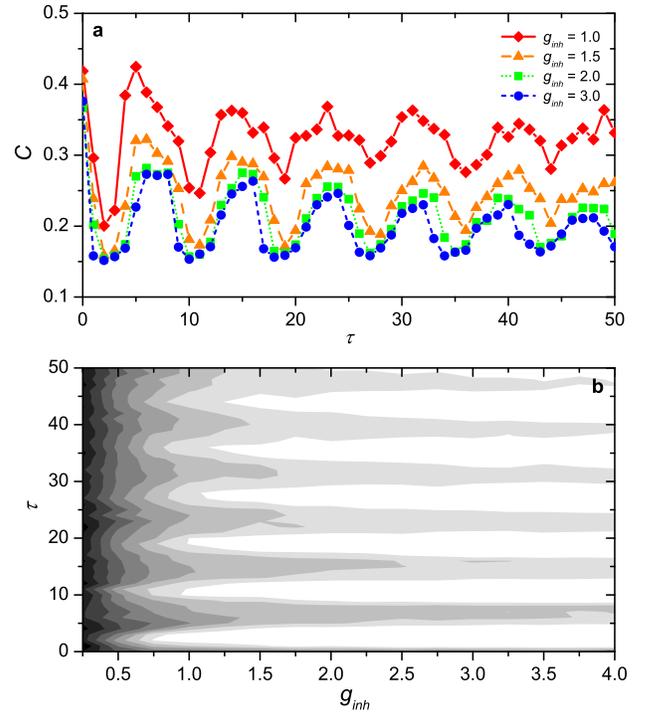}
\caption{\label{inhibcr}Delay-induced multiresonances in case of inhibitory coupling of the two neurons. Panel (a) shows the coherence measure $C$ in dependence on $\tau$ for different values of $g_{inh}$. As in Fig.~\ref{hybcr}, it holds that the stronger the coupling the better expressed the recurrently appearing minima of $C$. However, in the considered span of $\tau$ values, twice as many minima as by hybrid coupling can be observed. Panel (b) features the contours of $C$ (white depicts minimal and black maximal values) on the corresponding delay-strength $\tau-g_{inh}$ parameter plane, where the much denser tongues of coherency are clearly inferable. This indicates that the interplay between the synaptic delay $\tau$ and the characteristic time scale of the two Hodgkin--Huxley neurons is more efficient by purely inhibitory coupling. Other parameter values are: $\sigma=1.5$.}
\end{center}
\end{figure}

Turning to the second coupling scheme relying only on inhibitory synapses, however, we find somewhat unexpected results. While the time courses of the transmembrane potential $V$ presented in Fig.~\ref{inhibtime} do not suggest quantitatively different behavior in that certain values of $\tau$ warrant higher coherency of spiking than other values (which is also what we can observe in Fig.~\ref{hybtime}), a more accurate quantitative analysis presented in Fig.~\ref{inhibcr} indicates otherwise. In particular, in panels (a) and (b) we find that the minima of $C$ are much more frequent in the considered span of $\tau$ values as this was the case for hybrid coupling. While for the later a total of three minima can be observed within $0 \leq \tau \leq 50$ (see Fig.~\ref{hybcr}), for purely inhibitory coupling twice as many minima are inferable within the same span of $\tau$ values.

The origins of these multiresonant phenomena can be linked to different inherent properties of neuronal dynamics. First, it is useful to define the so called average excitatory time $T_e$, which is the average time between two consecutive spikes. For an isolated Hodgkin--Huxley neuron driven by noise this time decreases and saturates towards $T_e \approx 16$ for $\sigma \geq 4.0$ (note that this corresponds to a strong noise limit, above which the system may already exhibit numerical instability). Increasing the noise intensity further and lowering the time step for numerical integration, it is in principle possible to arrive at even lower average excitatory times $T_e \approx 12$, which agrees with the theoretical prediction stemming from the imaginary parts of the complex conjugate eigenvalues ${\rm Im}\lambda_{i,j} = \pm {\rm i}\omega = \pm {\rm i} 0.54$, where $T_e=2 \pi/\omega=11.63$. Since in our simulations, however, we use a comparatively low noise intensity $\sigma=1.5$, the average excitatory time $T_e \approx 16$ of an isolated Hodgkin--Huxley neuron is the more accurate approximation for the inherent time scale of the considered neuronal dynamics. For hybrid coupling, we thus find the first minimum of $C$ at $T_e/2$, and subsequent minima at odd multiples of the half of the average excitatory time [see Fig.~\ref{hybcr}(a)], which agrees with the doubly effect of the two considered synaptic types \cite{wang_po11}. The average excitatory time is reflected also in the time courses presented in Figs.~\ref{hybtime}(b,d,f) (note that in these the firing is accurate and ordered due to the constructive impact of $\tau$), where the average spiking period is approximately equal to $T_e$.

\begin{figure}
\begin{center} \includegraphics[width = 8.1732cm]{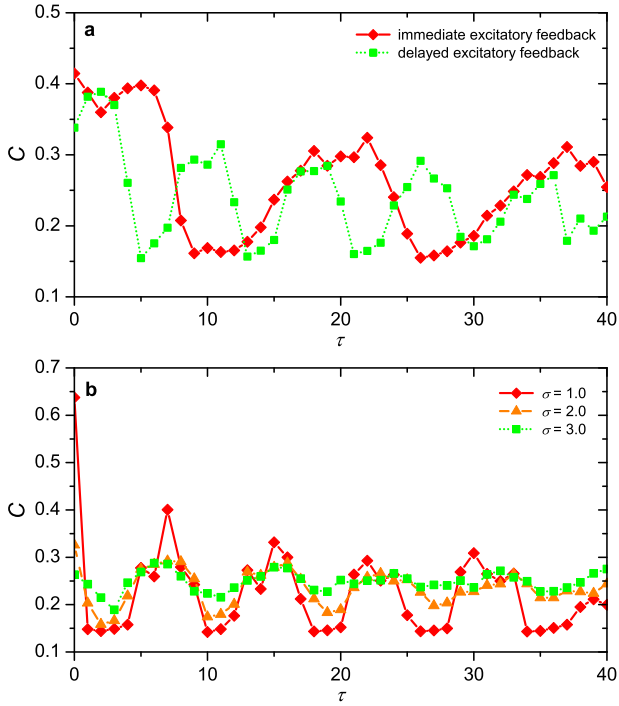}
\caption{\label{noise}Delay-induced multiresonances in the presence of additional delay in the excitatory feedback and noise. Panel (a) features a comparison of the coherence measure $C$ as obtained with and without excitatory synaptic delay in dependence on $\tau$ for hybrid coupling. It can be observed that the introduction of delays in the excitatory feedback can substantially reduce delays warranting the most coherent response. Other parameter values are: $g_{exc}=1.0$, $\sigma=1.5$. Panel (b) depicts $C$ in dependence on $\tau$ for different values of the noise intensity $\sigma$ in two purely inhibitory coupled neurons. It can be observed that as the intensity of noise increases the maximally attainable values of $C$ decrease (yet the effect saturates for higher $\sigma$). Optimal delays, however, remain unaffected by noise, which indicates robustness of the observed delay-induced multiresonances.}
\end{center}
\end{figure}

Conversely, for inhibitory coupling, the matching of the time scales leading to the multiresonant dependence of $C$ on $\tau$ is different. Although the average excitatory time $T_e \approx 16$ is likewise [as in Figs.~\ref{hybtime}(b,d,f)] reflected in the corresponding time courses presented in Figs.~\ref{inhibtime}(b,d,f), which have the same average inter-spike interval, twice as many minima imply that the resonant matching occurs not just for odd multiples of $T_e/2$, but in fact for odd and even multiples. However, all the minima of $C$ are preceded by a small delay of $2$s (where the first minimum occurs) that is necessary for the first resonant response. Since the purely inhibitory type of synaptic coupling lacks the excitatory input that is present by hybrid coupling, in the former case the matching of the time scales is twice as efficient.

\begin{figure}
\begin{center} \includegraphics[width = 8.30592cm]{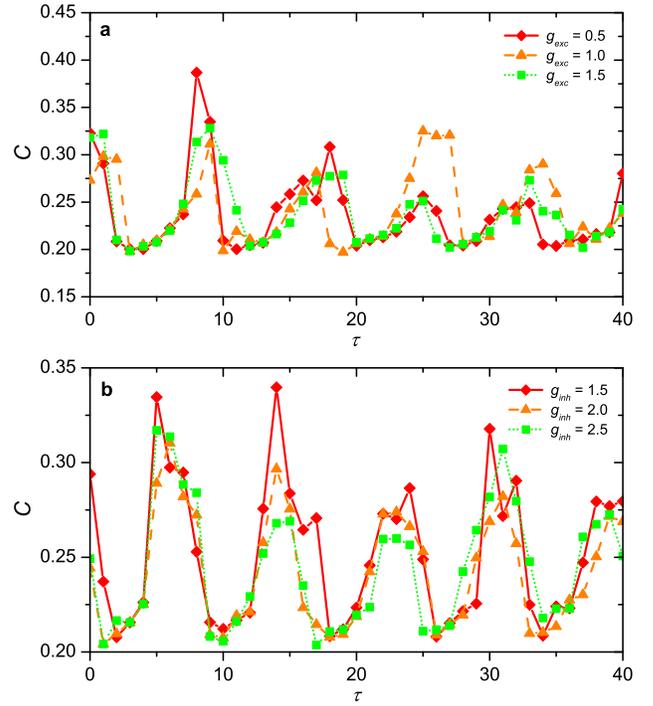}
\caption{\label{ring}Delay-induced multiresonances in a ring network consisting of $100$ neurons. Panel (a) features results as obtained with delay in the excitatory feedback ($g_{exc}$) and hybrid coupling. Panel (b), on the other hand, depicts $C$ in dependence on $\tau$ as obtained with delay in the inhibitory feedback ($g_{inh}$) and purely inhibitory coupling. Based on the presented results, it can be concluded that multiresonances in a ring network can be observed irrespective of the coupling and delay type, if only the delays are appropriately adjusted. However, delays warranting optimal coherence in the network with purely inhibitory coupling (b) are smaller that those in the network with hybrid coupling (a). Other parameter values are: $g_{inh}=1.5$ [applicable for panel (a) only] and $\sigma=1.5$.}
\end{center}
\end{figure}

Finally it is of interest to examine the robustness of our findings in the presence of delayed excitatory feedback, different levels of noise, and for different sizes of the network. In Fig.~\ref{noise}(a), we present the results with and without delayed excitatory feedback in a hybridly coupled two-neuron system. It can be observed that, while multiresonances can be observed in both cases, the introduction of delays also in the excitatory feedback (in addition to delays in the inhibitory feedback) may substantially reduce the delays that warrant an optimal response of the system (maximal values of $C$). Thus, delayed excitatory feedback does affect the results quantitatively, yet it does not affect the qualitative picture. Figure~\ref{noise}(b) shows that different noise intensities $\sigma$ have a similar impact. In particular, while higher values of $\sigma$ may reduce maximally attainable values of $C$, the multiple maxima are always clearly inferable and their positions do not shift. Hence, noise is also unable to significantly affect the results. Lastly, we present in Fig.~\ref{ring} results obtained on a larger ring network for the two different coupling types. Regardless of whether the coupling is hybrid with delays introduced to both types of synapses [panel (a)] or purely inhibitory [panel (b)], the multiple coherence resonances are clearly inferable. Importantly, also on larger networks the purely inhibitory mode of interneuronal communication appears to be more efficient (there are more maxima of $C$ in a given span of $\tau$) than the hybrid mode, which fully agrees with our conclusions obtained by means of the analysis of the two-neuron system, and thus solidifies the high robustness of our main conclusions, which we will summarize in what follows.

\section{Summary}
Summarizing, we have demonstrated the occurrence of multiresonant elevation of firing precision, as quantified by means of a coherence measure, in synaptically coupled Hodgkin--Huxley neurons. We have separately considered hybrid and purely inhibitory coupling, and we have discovered that the resonant matching of the different time scales that are inherent to the Hodgkin--Huxley model (and the information transmission delay) is twice as efficient in the latter case. Our results thus reveal unexpected possibilities for the resonant enhancement of firing precision by means of matching of different time scales of neuronal dynamics. Moreover, we have examined the robustness of our findings to the introduction of delay in the excitatory feedback, to noise, and to the number of coupled neurons. We have found that delayed excitatory feedback may substantially reduce the length of delays that ensure an optimal response of the system, yet that it does not qualitatively affect the results. Neither do noise and the size of the network, which led us to the conclusion that the reported results are highly robust, and that they are thus expected to remain valid also in other related neuronal systems. We hope that our study will prove useful for facilitating the development of concepts such as function-follow-form \cite{volman_pb05, zhou_prl06} and the application of methods of statistical physics for better understanding conditions such as epilepsy \cite{gao_pa05, lai_prl07, volman_po11} and other neurodegenerative diseases, as well as for better understanding the mechanisms behind high-precision firing patterns in more realistic neuronal networks.

\begin{acknowledgments}
This research was supported by the National Science Foundation of China
(grants 11172017 and 10832006) and by the Slovenian Research Agency (grant J1-4055).
\end{acknowledgments}

\end{document}